\documentclass[11pt]{article}
\usepackage{amsmath}
\usepackage{amsthm}
\usepackage{graphicx}
\usepackage{multicol} 
\usepackage{subfigure}
\usepackage{cite}
\usepackage{url}
\usepackage{color}
\usepackage{authblk}
\usepackage{amssymb}
\usepackage[margin=1.1in]{geometry}

\begin{document}

\title{Testing Wireless Sensor Networks with Hybrid Simulators}

\author[1,2]{Sain Saginbekov}
\author[2]{Chingiz Shakenov}
\affil[1]{Department of Computer Science, Nazarbayev University}
\affil[2]{National Laboratory Astana, Nazarbayev University}
\affil[1,2]{\textit {\{sain.saginbekov,chingiz.shakenov\}@nu.edu.kz}}

\renewcommand\Authands{ and }
 \date{}       
\maketitle

\begin{abstract}

Software development for Wireless Sensor Networks (WSNs) is challenging due to characteristics of sensor nodes and the environment they are deployed in. Testing software in a real WSN testbed allows users to get reliable test results. However, real testbeds become more expensive as the number of sensor nodes in the network grows. Simulation tools are alternatives to real testbeds. They are cheaper, faster and repeatable. However, simulation results are not reliable as that of testbeds. Therefore, there is a need for a testing tool that can leverage the advantages of testbeds and simulation tools. These tools are usually called hybrid simulators. In this survey, we discuss several hybrid simulators that use real sensor motes integrated with a simulator to make software development cheaper, repeatable and to make the results more reliable.   

\end{abstract}

\section{Introduction}
{\let\thefootnote\relax\footnote{This work has been funded by the Committee of Science of the Ministry of Education and Science of the Republic of Kazakhstan through the program N0.0662 ``Research and development of energy efficiency and energy saving, renewable  energy  and  environmental protection  for 2014 - 2016".}}

With the availability of cheap sensor nodes now it is possible to use hundreds of nodes in a Wireless Sensor Network (WSN) application. WSN applications have been being used in a wide range applications, including environmental, industrial, military, health-care and indoor applications~\cite{wsnsurvey}. WSNs are composed of sensor nodes, also known as \emph{motes}, equipped with a processing unit, a transceiver unit, a power unit and a
sensing unit. A typical sensor node usually is operated by battery power, has a limited memory and computation capability. These characteristics of sensor nodes restrict the node's capabilities. Moreover, WSNs could be deployed in different areas with different environmental conditions. For example, in~\cite{treedeployment} motes are deployed on trees while in~\cite{volcano,glacial,outofbox} motes were deployed in harsh and hazardous places. Due to the WSN characteristics and the environment they are deployed in, WSN applications may not work as expected and may live shorter than expected. Moreover, because of the above and other reasons, WSNs may experience different faults~\cite{lineinthesand}. Therefore, 
debugging and testing of applications written for WSNs is a difficult task at hand. 

Before installing and running a WSN application in a field, researchers need to check the protocol's correctness, scalability, energy efficiency, fault tolerance, etc. To do that researchers need a simulation tool or a testbed with real sensor motes. Simulating an application on a programming language without WSN models is the simplest, cheapest, but the most unreliable way of testing any WSN application as it does not need sensor nodes and does not consider different features of WSN networks such as link failures. To make the simulation more reliable a modeling of link qualities, hardware elements and other WSN features is required like in~\cite{tossim, avrora}. Furthermore, to make the WSN development more convenient, simulation codes should be able to be in installed in real motes. One attractive thing of a simulation, along with its price and speed, is that one can repeat running the simulation in different sized networks a number of times. However, simulations may not reveal errors that may occur in realistic conditions~\cite{simprob1}. For example, in~\cite{h-tossim}, unlike testbeds, simulations could not catch packet errors. Moreover, simulations cannot estimate the energy consumption of a particular application~\cite{h-tossim}.

Testing a WSN protocol on a testbed is more reliable as it runs on real sensor nodes and uses real radio links. During the testbed experimentation the correctness of the protocol can be checked and possible errors can be seen~\cite{h-tossim}, and the errors can be addressed before the deployment. However, it becomes expensive when it comes to testing the protocol's scalability and correctness in a large network. 


As mentioned above, the testing tools have their own strengths and weaknesses. Real testbeds are expensive whereas simulations are less realistic. Simulations could be fast and repeatable whereas testbeds may reveal problems that may occur in a real deployment.



Therefore, to benefit from the reliability of testbeds, repeatability, scalability and speed of simulations, and  to decrease the gap between simulation and testbed, another type of tool should be used. This kind of tool is called hybrid simulator, where real nodes and virtual nodes (nodes in the simulation) are used in parallel. In a hybrid simulation, along with a software simulation, at least one real sensor mote is used to make the simulation results more realistic. For instance, instead of using data and communication link models, if we use real sensor node data and communication links in our experiments, we could obtain more realistic results. However, unlike pure simulators, where simulation time does not necessarily need to be strictly correlated to actual time as long as correct sequencing is maintained, the hybrid approach requires coordination among virtual and real nodes~\cite{a-hybrid}.

There exist surveys on WSN simulators~\cite{sim1,sim2,sim3} and testbeds~\cite{test1,test2}. However, to our knowledge, there is no survey on hybrid simulators. Therefore, in this paper, we review exiting in the literature hybrid simulators that use virtual and real sensor nodes. 

\section{Hybrid Simulators}   


Although hybrid simulators differ in many ways, all of them integrate simulators with at least one real sensor node. And the purpose of using real sensor nodes may vary. For example, if some hybrid simulators use real sensor nodes for only getting data from sensors, some use real sensor nodes for using wireless links for communication. Most of the approaches discussed in this paper are mature enough and they have their advantages and disadvantages. Most of them are add-ons to existing WSN simulators, yet the others are dedicated to hybrid simulations from the start. In their comparison we are considered several essential characteristics, like scalability, synchronization between simulated environment and real testbed, GUI visualization, and emulation of hardware source code in simulated nodes (see Table~\ref{table1}). We don't give comparison according to the run-time performance or memory usage, because the goal of paper is to provide readers a broader view to existing hybrid simulators and used methods. In this section, we discuss existing in the literature hybrid simulators in chronological order.

\subsection{SensorSim}

In order to model power usage, allow the interaction between virtual and physical nodes, and enable real time user interaction with graphical data display, the authors of~\cite{sensorsim} proposed a simulation framework called \emph{Sensorsim}. Sensorsim extends the ns-2 simulator~\cite{ns2} in a sensor network context by providing new a power model, a new GUI, and supports hybrid simulations~\cite{sensorsim}. The power model simulates the hardware (CPU, radio module and available sensors). The main purpose of interaction with real nodes in Sensorsim is to get a good quality of sensor measurements. However, as the simulation scheduler and protocol execution in real nodes are not synchronized, real nodes cannot interact with virtual nodes at the correct time.

\subsection{MULE}  
 

A hybrid simulator proposed in~\cite{mule}, called MULE, eases the debugging stage of software development with multiple simulated motes 
on a host PC, i.e., by making it possible to inspect the state of the simulated motes. 
This also means that entire application can be suspended and restarted during the 
execution phase as needed. By carrying out the radio transmission and sensor readings 
with actual physical motes MULE is able to improve the fidelity of message transmission 
and sensor data acquisition. MULE is based on TOSSIM, a TinyOS simulator.

MULE runs the application on the host PC with TOSSIM but when it comes to communication
and sensing it transfers these to the physical motes connected to the PC. I.e., the actual
components of TOSSIM responsible for simulation of communication and sensing are replaced
with components that handle interaction with physical motes.

The coordination of simulated time of the motes on a PC and the actual time of transmission 
and data sensing is handled by MULE, i.e., whenever a real event is requested MULE first 
suspends/freezes the simulation by keeping the record of simulated
timing parameters for this specific event. Next it translates these parameters into real
time and performs the execution of the event in real time using physical motes and gather
timing parameters. Finally, these parameters are translated back to simulated time and the
execution of the simulated application resumes.

Although empirical study was performed only with upto 5 motes in~\cite{mule} and there were future
works left to be considered such as component migration, tiling, and time synchronization
to be added to MULE. These were recently incorporated into MULE in~\cite{msthesis} with special attention
to scale, fidelity, and speed. Empirical results for simulated environment were presented 
upto 10,000 motes, and percentage of fidelity was shown to be around 80-85 
physical motes.

\subsection{EmStar}

EmStar~\cite{emstar} is a framework targeted at more powerful nodes called \emph{Microservers} such as Crossbow's Stargate platform. However, it can also be used for less powerful motes. EmStar is composed of small reusable components which makes it possible to plug in new application-specific components. Other than running pure simulations, EmStar can run hybrid simulations using real radio for communication. EmStar has a range of tools to support the simulation. The tools run the simulation by modeling radio and sensor channels, provide an interface to a real radio (in hybrid simulations), provide graphical visualization and manage running services. Also, EmStar provides several services. The functions of these services include monitoring and providing link qualities, different types of routing and time synchronization among nodes. The code written for EmStar can also be installed into real motes like in TOSSIM~\cite{tossim}.   
Nevertheless, EmStar is more suitable for more powerful platforms. Therefore, in~\cite{sim-emu-depl}, the authors extended EmStar and proposed tools that support simulating a heterogeneous network, a network of less and more powerful nodes, with measurement and visualization. The tools also support hybrid simulations like EmStar. 

\subsection{SEMU}

To ease study the behaviour of the specific application and protocol and help in understanding this behaviour before the actual deployment a framework of simulation environment (\emph{SEMU}) for WSN with co-simulation, hybrid, model was presented in~\cite{Lo2007}. The co-simulation model proposed in SEMU is similar to the three hybrid modes for WSN simulation as of that proposed for \emph{EmStar}~\cite{emstar}. However, SEMU is able to provide development of WSN application with more accurate model by introducing the possibility of run time profiling to resolve synchronization problems.

The main distinction of SEMU from other hybrid simulators is that it uses virtual machines (VM) to mode the simulation in a network. In an abstract level the framework of SEMU consists of five layers. From top-to-bottom view the framework consists of five layers, these are VM, Communication, virtual OS, module, and native OS layers. The top most layer, VM, is where the virtual nodes, real nodes, and distributed GUI reside. A virtual node is represented as an emulation of a real node, and can consist of several VMs. Virtual nodes interact with the simulation environment through the communication links between VM and communication layers. So the purpose of communication layer is to serve as a bridge enabling communication between VM and virtual OS layers, where the latter acts as the control center of the simulation. (One of the components of this layer, called SIM Kernel, acts as a service provider by helping virtual nodes to forward back and forth their service requests into module and native OS layers.) Next comes the module layer, which provides components/tools to enhance the functionalities and management of the simulation framework. Briefly, these include maintaining all simulation status and references of the simulation objects, analyzes of log information with classification, scheduler for simulation of multiple nodes executing in parallel, time manager for resolving synchronization problems, abilities to configure devices and integrate custom protocols, and several models are provided to reflect the real conditions in a physical environment.

The co-simulation model of SEMU supports the real communication and sensing channel by utilizing the real physical node devices (an agent). In this model SIM Kernel serves all of the virtual nodes to request actions, such as sending, receiving and sensing. The process is as follows:
\begin{enumerate}
\item SIM Kernel chooses an agent, which is capable of representing the realistic behaviour of the node;
\item Agent is chosen for request: agent performs the corresponding action and provides run time profiling information;
\item Action is accomplished: agent notifies SIM Kernel of results, which also include execution time and the status for the action;
\item Based on results gathered so far SIM Kernel checks with time manager whether the action could be completed;
\begin{enumerate}
\item Request is granted: SIM Kernel return all the data to the virtual node;
\item Request is denied: SIM Kernel blocks the virtual node until the request is granted by time manager.
\end{enumerate}
\end{enumerate}

According to this co-simulation model it is important to track the cooperation between the virtual and real physical nodes (agent). I.e., virtual nodes should take into account the execution time of the agents and make sure they access the data in accordance to their virtual simulation time. Hence this co-simulation model needs to provide the synchronization of interaction, time and data. Once the action requested by the virtual node is accomplished by the agent, it will return the status and real execution time of the action. This makes sure that interaction is synchronized. Next, data synchronization is achieved through the help of the environment recorder, which collects information from physical environment in wall time. The conversion of wall time to virtual time is performed by the environment recorder. Environment recorder keeps all of these information in a the storage. Finally, the time synchronization is achieved by profiling the execution time of action in an agent, these results are represented as number of clock cycles. Thus, virtual time is derived from translation of elapsed clock cycles according to cycles per instruction (CPI). The virtual time helps the time manager in deciding whether to grant an action or not.

For experimental results the virtual nodes were setup in a grid and the event was broadcast from the corner node. Then the event was distributed using the flooding protocol until all of the virtual nodes received it. It was reported that SEMU did run fast below 1250 virtual nodes~\cite{Lo2007}. When the number of virtual nodes is increased beyond 1250 it was reported that simulation time quickly increases due to the limited RAM that the simulation was carried on, which was 1.5 GB \cite{Lo2007}. Additionally, authors of~\cite{Lo2007} stated that scalability of SEMU is highly affected by the resource management of the Linux OS (i.e., the native OS).


\subsection{Simulation-Based Augmented Reality}  

Another WSN development scheme, that uses both virtual and real nodes to make the simulation scalable, flexible, controllable and make WSN development easier, has been proposed in~\cite{augmented}. The idea is to design a network of real and virtual nodes that interact with each other in a seamless and transparent manner. For that nodes called ``super nodes" are used to bridge network traffic between virtual and real nodes and synchronize virtual and real time. Super nodes are real nodes that are connected to virtual nodes via a serial interface. Only a subset of real nodes could be selected as super nodes. The selection of super nodes depends on the network topology. Virtual nodes that are in the radio range of real nodes should be super nodes as they should be able to communicate with the real nodes. This requirement, however, may demand many real nodes if the simulated network is dense.




\subsection{Sensornet Checkpointing}

Another hybrid testing tool has been proposed in~\cite{checkpoint}. However, their approach differs from others in that it allows testbed and simulation network checkpointing. That is, nodes in the network could store the states of certain elements such as memory, timer system, radio, etc., and start from any of the checkpointed state again. Therefore, along with the realism, the proposed approach allows researchers to make tests repeatable unlike  pure testbeds and hybrid testbeds proposed by others. Moreover, the approach benefit from the advantages of simulations like injecting faults, visualization and debugging testbeds.

Before getting or setting states, a network is send a single byte command to freeze all nodes. Then, after getting or setting states, another command is send to unfreeze the network. All nodes in the network freeze and unfreeze at the same time. After getting network states from the testbed, one can visualize the network in simulations, simulate errors by changing a state and moving back the modified state to the testbed, and can repeat the experiment from any network state.


However, as every node is connected to a virtual node in the approach, it is difficult to test large networks with this approach.

\subsection{ATOSSIM}

In~\cite{a-hybrid}, the authors propose a hybrid simulation framework, called ATOSSIM, in order to provide reliable simulation. ATOSSIM is an augmented version of the TOSSIM simulator. ATOSSIM enables virtual and physical nodes work together seamlessly. Unlike TOSSIM where all nodes run the same code, ATOSSIM allows to run different codes on specialized nodes called \emph{shadow} nodes. However, the main purpose of shadow nodes is to bridge virtual nodes with real nodes. ATOSSIM uses an application gateway to ensure correct coordination among virtual, shadow and real nodes. The application gateway does that by artificially adding a delay which is computed as the difference between the time when the event is selected by the scheduler and the time stored in the event handler descriptor~\cite{a-hybrid}. 

The experiment results show that timing error increases with the number of simulated nodes, and the error becomes noticeable when the number of simulated nodes is higher than $50$. Therefore, ATOSSIM results may not be reliable for large networks.

\subsection{H-TOSSIM}
One of the widely used WSN simulation tools is TOSSIM~\cite{tossim}. TOSSIM is designed for TinyOS~\cite{tinyos}. However, as mentioned earlier, simulation tools such as TOSSIM are not realistic. With TOSSIM it is difficult to detect faults that may occur in real settings, for example, faults that occur due to timing, communication links, task calculation overload and message length setting error. Moreover, TOSSIM does not support the power consumption estimation of sensor nodes. Therefore, in~\cite{h-tossim}, the authors extend TOSSIM, and calls it H-TOSSIM, to make TOSSIM more realistic. The authors propose a hybrid testbed that combines virtual nodes with physical nodes. A combination of virtual nodes with physical nodes benefits from the advantages of simulation and testbed experiments. The experiments conducted with H-TOSSIM showed that H-TOSSIM reveals program errors which simulations could not capture. Moreover, the experiments showed that H-TOSSIM supports power consumption estimation for large WSN with high accuracy and low hardware cost. 

H-TOSSIM uses three physical nodes. One of the physical nodes shares the simulated environment with virtual nodes, other two are used to bridge the physical node with the computer. There are two reasons of using two physical nodes for bridging: i) to make it possible to send concurrent messages from the computer to the physical node and ii) to match the rate of radio and serial links as the rate of radio can be up to $250 Kbps$ whereas the serial link rate can be at most $115 Kbps$~\cite{h-tossim}.

To synchronize virtual time and real time, H-TOSSIM checks if the virtual time is faster than the real one, if it is faster, H-TOSSIM goes to sleep until it reaches the appropriate time (when virtual and real time will be equal), otherwise it executes immediately.
  
\subsection{Simulating Protocols on Large Networks}

\begin{figure*}[!h]
\vspace{-5mm}
\centering
\subfigure[Source node $A$ sends to relay node $B$]{
\includegraphics[scale=0.27]{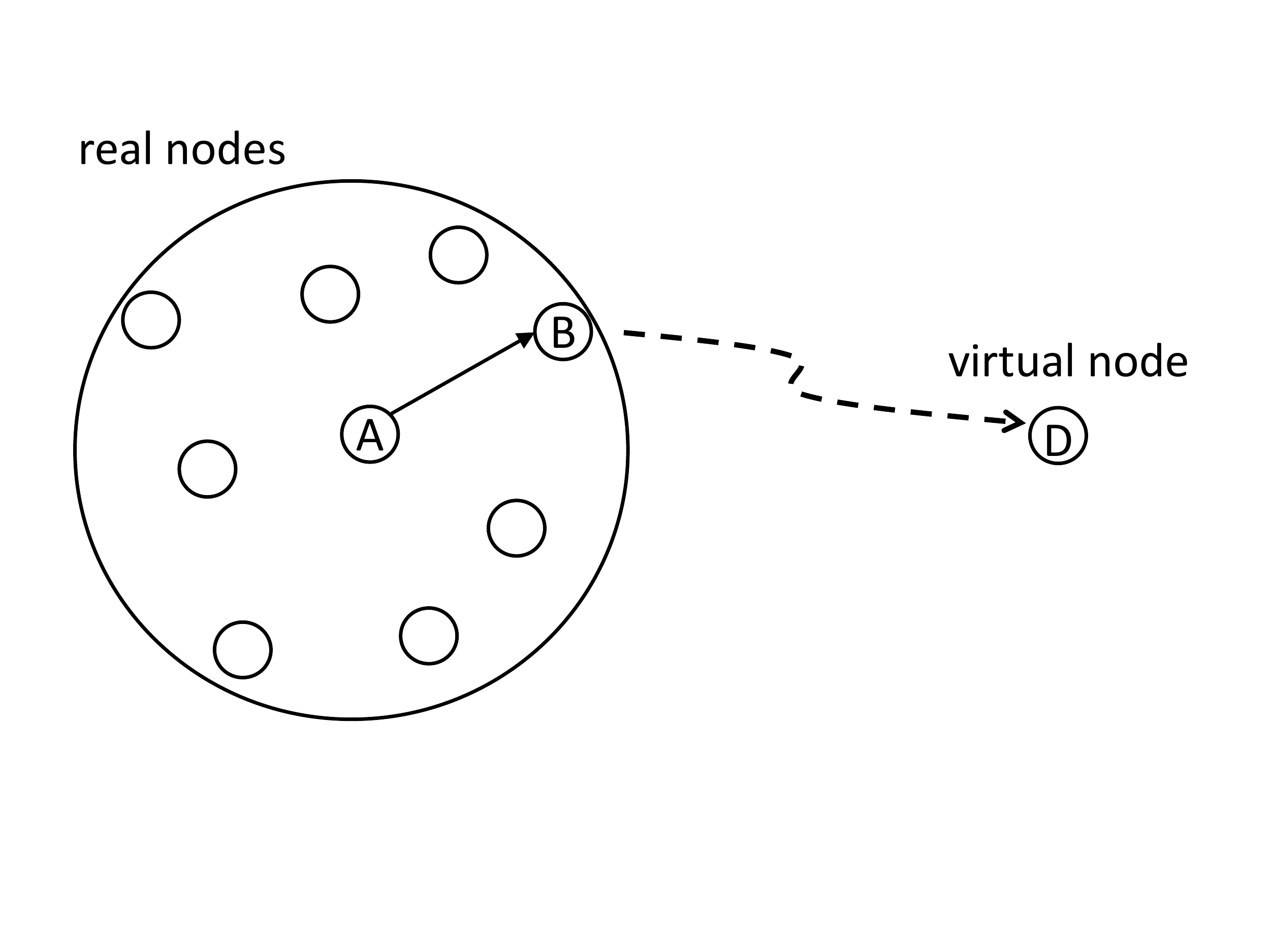} 
\label{fig:orig}
}
\subfigure[$A$ becomes $B$ and sends to the next relay node $C$ toward virtual destination]{
\includegraphics[scale=0.27]{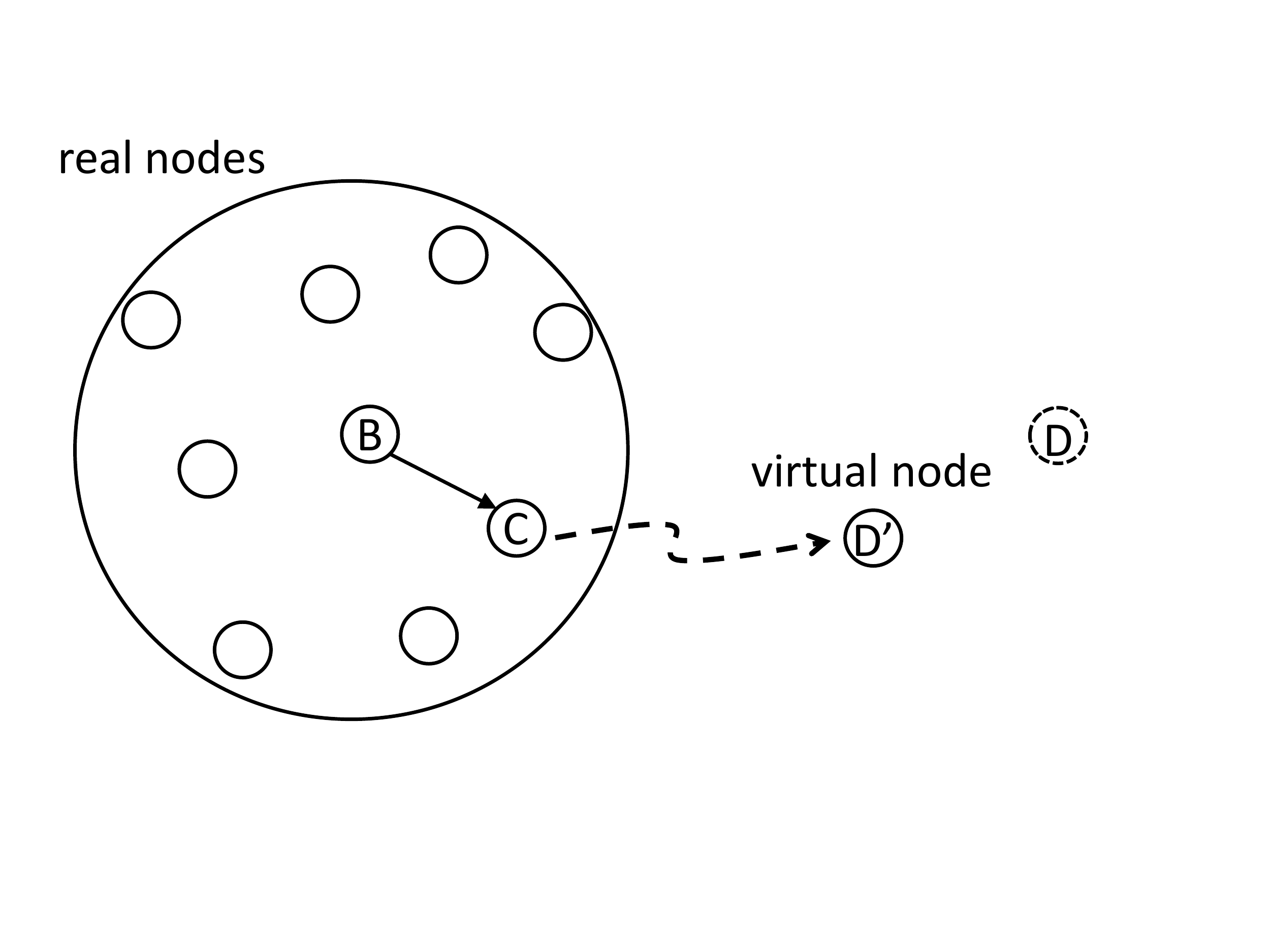} 
\label{fig:next}
}
\vspace{-4mm}
\caption[]{Greedy Routing example in a network with real nodes and virtual nodes combined}
\label{fig:greedy1}
\end{figure*}

The paper proposes a new way of simulating protocols on a large network with only small number of real nodes by combining real nodes with virtual nodes. The authors use this method to simulate greedy geographic routing algorithm~\cite{greedy1}. The idea is to have a network consisting of real nodes and a virtual node, where the source node and its one-hop neighbours are real nodes (placed randomly), while the destination node is the virtual node. When the source node $A$ selects one of its neighbouring node, $B$, as a relay node (according to the routing algorithm) and sends a message to $B$, the location of the destination node changes with the vector $\overrightarrow{-AB}$. This allows to make the source node $A$ as $B$, i.e., node $A$ sends the message again, but now $A$ sends as $B$ (see Figure~\ref{fig:greedy1}). This process is repeated until the virtual destination can be reached directly from the relay node. In this step, the relay node sends the message to the node which is the nearest to the destination. Note that the communication is done over the real radio link.

However, this approach does not simulate the real network as the neighbourhood consisting of real nodes is always the same, i.e., relay nodes will always have the same neighbours. Furthermore, the neighbourhood density will always be the same.

\subsection{From Real Neighbours to Imaginary Destination}

\begin{figure*}[!h]
\centering
\subfigure[After $B$ sends to relay node $C$, the location of $D`$ is rotated]{
\includegraphics[scale=0.27]{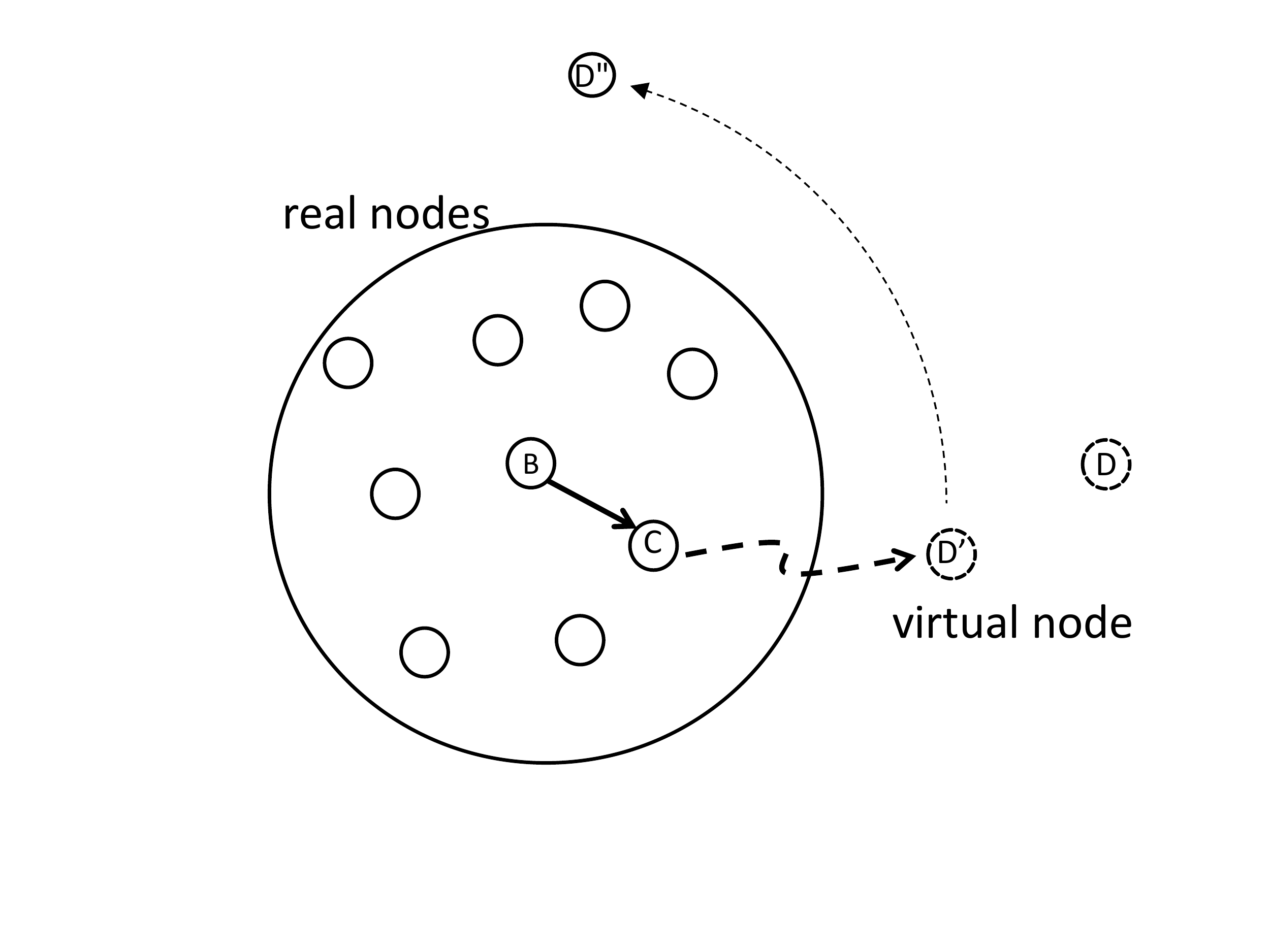} 
\label{fig:orig2}
}
\subfigure[$C$ sends to next relay node $E$ toward virtual destination]{
\includegraphics[scale=0.27]{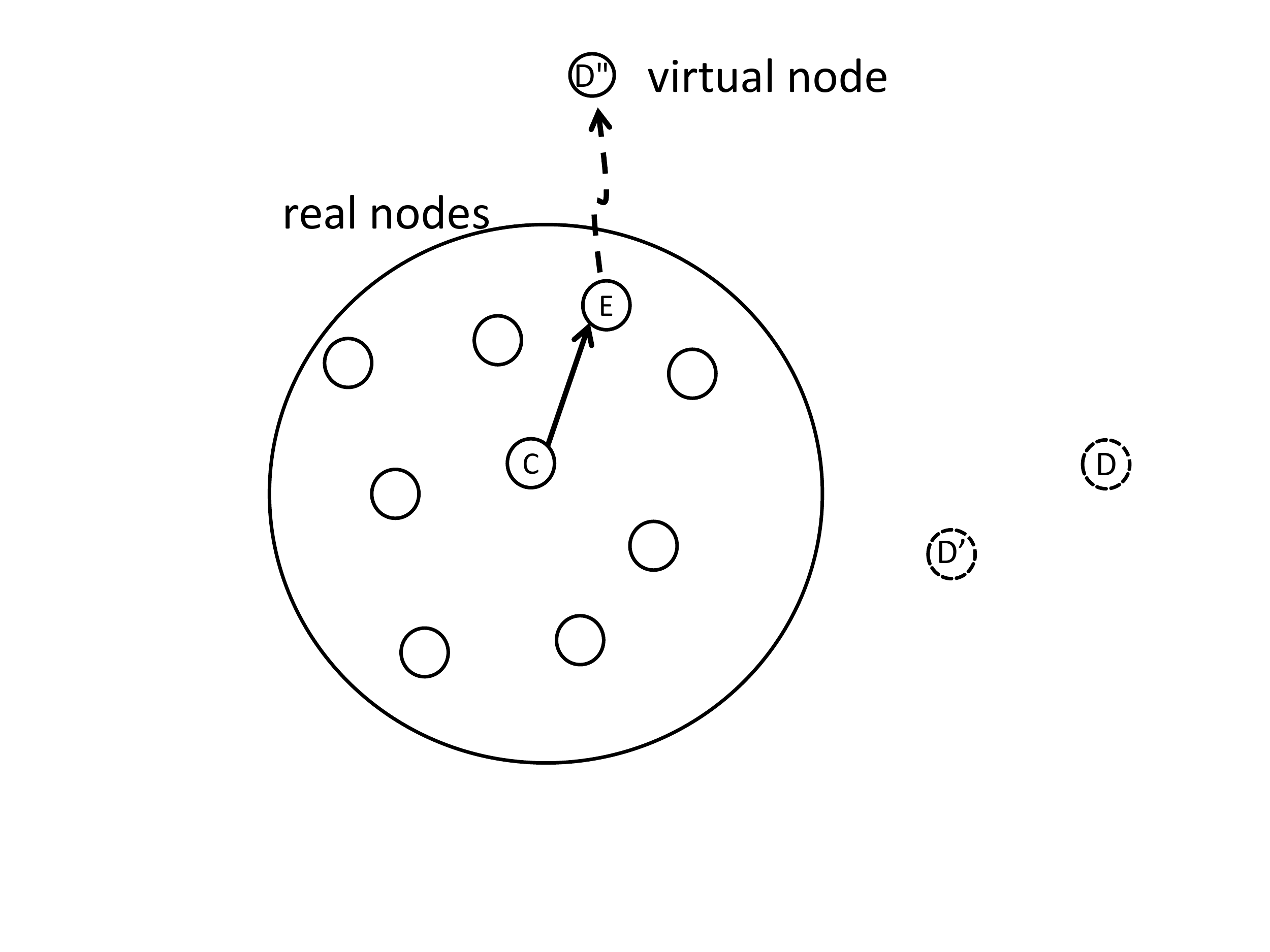} 
\label{fig:next2}
}
\vspace{-3mm}
\caption[]{Rotated virtual destination}
\label{fig:greedyrotate}
\end{figure*}

\begin{figure*}[!h]
\centering
\includegraphics[scale=0.27]{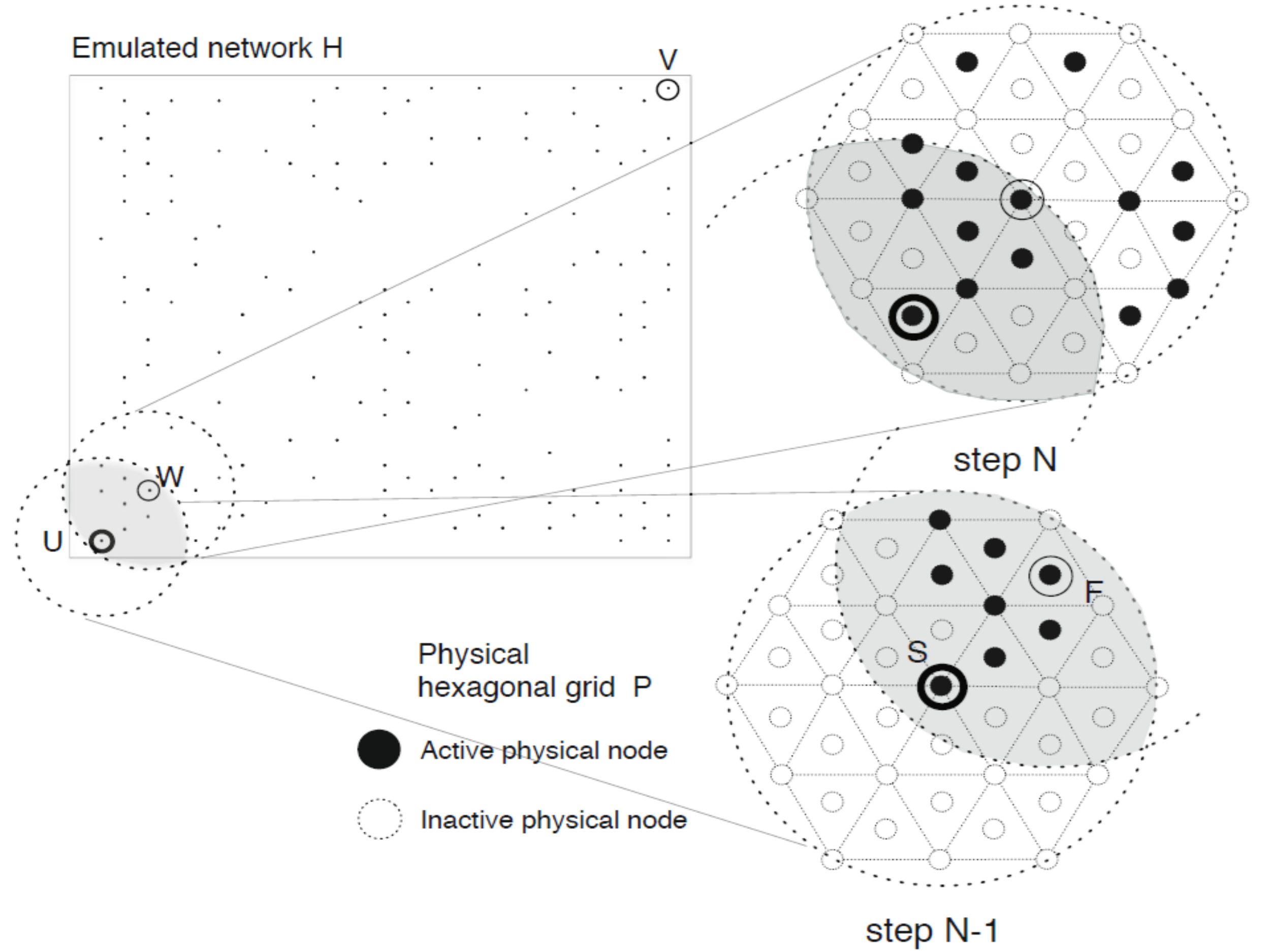}
\vspace{-3mm}
\caption{Mapping simulated nodes to real nodes~\cite{greedy2}}
\label{fig:hexagon}
\end{figure*}

To partially overcome the problem of the ``same neighbourhood" existing in~\cite{greedy1}, the destination node could be rotated around the source node (see Figure~\ref{fig:greedyrotate} for illustration). However, rotating the destination will not give full randomness~\cite{greedy2}. Therefore, the authors of~\cite{greedy2} proposed an improved version over~\cite{greedy1}. The idea is, instead of placing real nodes randomly, to place real nodes in a hexagonal grid, and map virtual nodes generated randomly (or other way) to them. A virtual node is mapped to a real node by rounding its coordinates to the hexagonal grid. At each step, the virtual neighbours of the source are mapped to the real nodes and only those real nodes become activated. After the source $A$ sends a message to the next relay node $B$, unlike in
~\cite{greedy1}, where only the virtual destination is translated, the entire emulated network is translated by vector $\overrightarrow{-AB}$ (see Figure~\ref{fig:hexagon} for illustration). Therefore, this approach supports the randomness and density of the emulated network. By placing appropriate number of real nodes in 1-hop neighbourhood, this approach can emulate large networks with varying densities. However, a limitation of the approach is if the emulated network is very dense, more than one virtual nodes could be mapped to one real node. 
                         

\vspace{3mm}

\subsection{Hy-Sim}
  
In~\cite{hysim}, the authors propose a model-based hybrid simulation framework, named Hy-Sim, for wireless sensor networks. 

A development flow in Hy-Sim consists four steps. In the first step, application level software is developed by MathWorks tools~\cite{mathworks} which provide mechanisms such as efficient code generation and platform-specific re-targeting~\cite{hysim}. In the second step, communication and sensor reading models can be configured to run in a fully simulated environment. In the third step, virtual nodes could be interfaced with real nodes to get real sensor data or communicate over real radio. The fourth step includes generating codes for real nodes for different WSN platforms.          

The Hy-Sim architecture consists of 3 blocks: Nodes block, Super-medium block and Super-sensor block. Nodes block contains each node instance that could be fully, partly simulated or independent. Super-medium block manages all radio communications within the simulated world and between simulated and real world. Super-sensor block manages all sensing aspects, provides access to actual sensors and simulated sensor readings. 		

In Hy-Sim, a node in the network could be one of the five types: SIM, HIL$\_$SEN, HIL$\_$RF, HIL$\_$FULL or REAL. A node is of type SIM if it is fully simulated, i.e., sensor readings and radio channels are simulated. Sensor readings and radio channels (link-quality) are defined in a file during the configuration phase. A node is of type HIL$\_$SEN if it is simulated but uses real sensor data received from real nodes. A node is of type HIL$\_$RF if it is simulated but communicates over real radio. A node is of type HIL$\_$FULL if it uses real sensor data and communicates over real radio.
And a node is of type REAL if it is a real sensor node.

\subsection{Comparison of hybrid simulators}
\begin{itemize}
\item Scalability

Usually, routing and network algorithms that are tested on small amount of motes works in a different way on large amount of motes. Ability to scale the simulation to big number of virtual motes and mirror it to real motes is considered to be a good advantage. Most of the proposed methods above implement this feature.
\item Synchronization

Running simulation in sync with real hardware enables to see the real difference of behavior between the simulated nodes and the real nodes at runtime. Unfortunately, sometimes, little time gap between virtual and physical events occurs in synchronized systems, caused by transferring state of virtual node to real physical node or vise-versa.
This problem was solved in H-TOSSIM and Sensornet Checkpointing simulation approaches, by freezing virtual environments and physical node states, where total time of freezing command is comparatively much smaller than total time of state transfer. Although accuracy of simulation increases, simulation time lasts longer.
\item GUI

GUI gives us a vision of processes of simulation. Even though every step of simulation can be logged to a file, it is harder to track concurrent events, especially when two or more nodes are working independently. It is also gives a good vision when simulation runs under real-time speed.
\item Source emulation

One of the most important feature of network simulation platforms is the source code emulation. It makes easier to develop algorithms, considering that the same code can be executed both virtually and physically. Otherwise, developers can't be sure that the simulated code act similarly (if not the same way) as a real device code. In different platforms it was realized in different ways.

Methods used in Mule, ATOSSIM and H-TOSSIM are inherited their functionalities from TOSSIM, simulator of TinyOS, as Sensornet Checkpointing method is based on Contiki-OS simulator Cooja. These simulators emulate source code by dynamically compiling it on simulation runtime.

Simulation on large network [16] and approach of hexagon formation [22] are based on Worldsens environment, which is consist of WSNet network simulator and WSim hardware simulator. WSim is used to debug the application using the real target binary code.

Hy-Sim is model based and can generate application code for hardware from the developed simulation model.
\end{itemize}

 \begin{table}[!h]
\begin{center}
  \begin{tabular}{| l | c | c | c | c |}
    \hline
    ~							& Scalable		& Synchronized	& GUI			& Source emulation	\\ \hline
    Sensorsim~\cite{sensorsim}	& ~				& ~				& \checkmark	& ~					\\ \hline
    Mule~\cite{mule}			& \checkmark	& \checkmark	& ~				& \checkmark		\\ \hline
    EmStar~\cite{emstar}		& ~				& \checkmark	& \checkmark	& ~					\\ \hline
    Semu~\cite{Lo2007}			& \checkmark	& \checkmark	& \checkmark	& ~					\\ \hline
    \cite{augmented}			& \checkmark	& \checkmark	& ~				& ~					\\ \hline
    \cite{checkpoint}			& \checkmark	& \checkmark	& ~				& \checkmark		\\ \hline
    ATOSSIM~\cite{a-hybrid}		& ~				& ~				& \checkmark	& \checkmark		\\ \hline
    H-TOSSIM~\cite{h-tossim}	& \checkmark	& \checkmark	& \checkmark	& \checkmark		\\ \hline
    \cite{greedy1}				& \checkmark	& ~				& ~				& \checkmark		\\ \hline
    \cite{greedy2}				& \checkmark	& ~				& ~				& \checkmark		\\ \hline
    Hy-Sim~\cite{hysim}			& \checkmark	& ~				& ~				& \checkmark		\\ \hline
 
  \end{tabular}
  
\end{center}
\vspace{-4mm}
 \caption{Hybrid Simulators}
\vspace{-1mm}
\label{table1} 
 \end{table}

\vspace{-8mm}

\section{Conclusion}

A number of experiments should be conducted before deploying a WSN in a region to check the correctness of software. There exit different simulators and testbeds that are used to test WSN software. Simulators are faster, repeatable and cheaper, whereas testbed results are reliable. A hybrid simulator integrates a simulator with at least one real sensor mote.

In this paper, we discussed hybrid simulators developed for WSNs. Hybrid simulators integrate simulation tools and real sensor nodes to leverage the advantages of testbeds and simulation tools. On one hand, hybrid simulators, like simulation tools, are not expensive, repeatable, and usually scalable. On the other hand, hybrid simulators' results could be reliable as testbed results.

\bibliography{Surv_arXiv}{}
\bibliographystyle{abbrv}

\end{document}